\documentclass[prl,twocolumn,aps]{revtex4}

\usepackage{graphicx} 
\usepackage{color} 
\usepackage{rotating}
\usepackage{epsfig}

\newcommand{\ket}[1]{\left|#1\right\rangle}
\newcommand{\bra}[1]{\left\langle#1\right|}

\begin{document}

\title{Surface code quantum communication}

\author{Austin G. Fowler$^{1}$, David S. Wang$^{1}$, Charles D. Hill$^{1}$, Thaddeus D. Ladd$^{2,3}$, Rodney Van Meter$^{4}$, Lloyd C. L. Hollenberg$^{1}$}
\affiliation{$^{1}$Centre for Quantum Computer
Technology, University of Melbourne, Victoria, Australia\\
$^{2}$Edward L. Ginzton Laboratory, Stanford University, Stanford,
CA, 94305-4088, USA\\
$^{3}$National Institute of Informatics, 2-1-2 Hitotsubashi,
Chiyoda-ku, Tokyo-to 101-8430, Japan\\
$^{4}$Faculty of Environment and Information Studies, Keio
University, 5322 Endo, Fujisawa, Kanagawa, 252-8520, Japan}

\date{\today}

\begin{abstract}
Quantum communication typically involves a linear chain of repeater
stations, each capable of reliable local quantum computation and
connected to their nearest neighbors by unreliable communication
links. The communication rate in existing protocols is low as
two-way classical communication is used. We show that, if Bell pairs
are generated between neighboring stations with a probability of
heralded success greater than 0.65 and fidelity greater than 0.96,
two-way classical communication can be entirely avoided and quantum
information can be sent over arbitrary distances with arbitrarily
low error at a rate limited only by the local gate speed. The number
of qubits per repeater scales logarithmically with the communication
distance. If the probability of heralded success is less than 0.65
and Bell pairs between neighboring stations with fidelity no less
than 0.92 are generated only every $T_B$ seconds, the logarithmic
resource scaling remains and the communication rate through $N$
links is proportional to $(T_B\log^2 N)^{-1}$.
\end{abstract}

\maketitle

Long-range communication of quantum states is difficult as such
states cannot be copied \cite{Diek82,Woot82}. Current research into
long-range quantum communication focuses on quantum repeaters
\cite{Brie98} making use of entanglement purification
\cite{Benn96bs} and entanglement swapping \cite{Zuko93,Sang09}.
Entanglement purification requires slow two-way classical
communication, resulting in the quantum communication rate
decreasing polynomially with distance. Furthermore, the
communication error rate $p_c$ is at best comparable to the error
rate $p_g$ of gates within repeaters. If qubits have a finite
coherence time, requesting a constant $p_c$ as the distance
increases results in a finite maximum communication distance.
Arbitrarily rapid and reliable communication over arbitrary
distances is not possible using only entanglement purification and
swapping.

Initial work incorporating error correction into quantum
communication resulted in non-fault-tolerant schemes
\cite{Pers08,Pers09} capable of reliably correcting only a small,
fixed number of errors. Recently, the first steps towards
fault-tolerant quantum communication were taken \cite{Jian09},
however entanglement purification was still used between neighboring
quantum repeaters, fundamentally limiting the communication rate to
hundreds of logical qubits per second. A quantum communication
protocol requiring very little two-way classical communication has
been developed concurrent with this work \cite{Munr09}

We show that, using surface code quantum error correction
\cite{Brav98,Denn02,Fowl08,Wang09}, two-way classical communication
can be avoided entirely provided we can create Bell pairs between
neighboring stations with a heralded success probability $S_B\gtrsim
0.65$ and fidelity $F\gtrsim 0.96$. This means communication can
proceed at a rate independent of the classical communication time
between repeater stations. Given local quantum gates with $p_g\ll
0.75\%$, we show that it is possible to communicate logical qubits
over arbitrary distances with arbitrarily low $p_c$ at a rate
limited only by the local gate speed. The number of qubits per
repeater increases only logarithmically and the quantum
communication rate decreases only logarithmically with communication
distance.

To describe our quantum communication protocol, we must first
describe surface codes and this in turn requires the notion of
stabilizers \cite{Gott97}. A stabilizer of $|\Psi\rangle$ is an
operator $M$ such that $M|\Psi\rangle = |\Psi\rangle$. For example,
$Z|0\rangle = |0\rangle$. Given any set of commuting operators
$\{M_i\}$, a state $|\Psi\rangle$ exists stabilized by $\{M_i\}$.

\begin{figure}
\begin{center}
\resizebox{60mm}{!}{\includegraphics{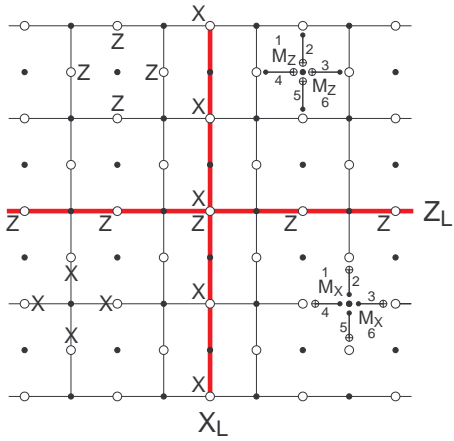}}
\end{center}
\caption{A surface code logical qubit. Stabilizers $ZZZZ$/$XXXX$ are
associated with the data qubits (open circles) around each
face/vertex. Syndrome qubits (dots) measure stabilizers using the
indicated sequences of gates. Logical operators $Z_L$, $X_L$ connect
opposing boundaries.}\label{fig:surface_code}
\end{figure}

Surface codes can be defined on lattices of the form shown in
Fig.~\ref{fig:surface_code}. Data qubits are represented by open
circles. We define a set of commuting operators on data qubits by
associating $ZZZZ$/$XXXX$ with each face/vertex. If the
$|\Psi\rangle$ stabilized by these operators suffers errors,
becoming $|\overline{\Psi}\rangle$, then local to these errors we
obtain equations of the form $M|\overline{\Psi}\rangle =
-|\overline{\Psi}\rangle$. Measuring whether the qubits are in the
$+1$ or $-1$ eigenstate of each stabilizer thus gives us information
about the errors in the lattice. Measuring a stabilizer requires a
sequence of six gates. This information can be used to reliably
correct the errors provided the error rates of initialization, CNOT,
measurement, and memory, which here we take to be equal at rate
$p_g$, are all less than approximately 0.75\%
\cite{Raus07,Fowl08,Wang09}. Logical operators $X_L$/$Z_L$ are
chains of single-qubit $X$/$Z$ operators that commute with every
$Z$/$X$ stabilizer and link the top/left boundary to the
bottom/right. The distance $d$ of the code is the number of
single-qubit operators in the shortest logical operator.

Transmitting surface code logical qubits is of particular interest
as the surface code possesses a high threshold error rate, requires
only local interactions, is highly tolerant of defective qubits
\cite{Stac09} and permits fast, arbitrarily long-range logical CNOT
--- a collection of properties no other scheme currently possesses. There are
a number of proposed architectures well-suited to implementing the
surface code \cite{VanM09,DiVi09,Amin09s}.

\begin{figure}
\begin{center}
\resizebox{85mm}{!}{\includegraphics{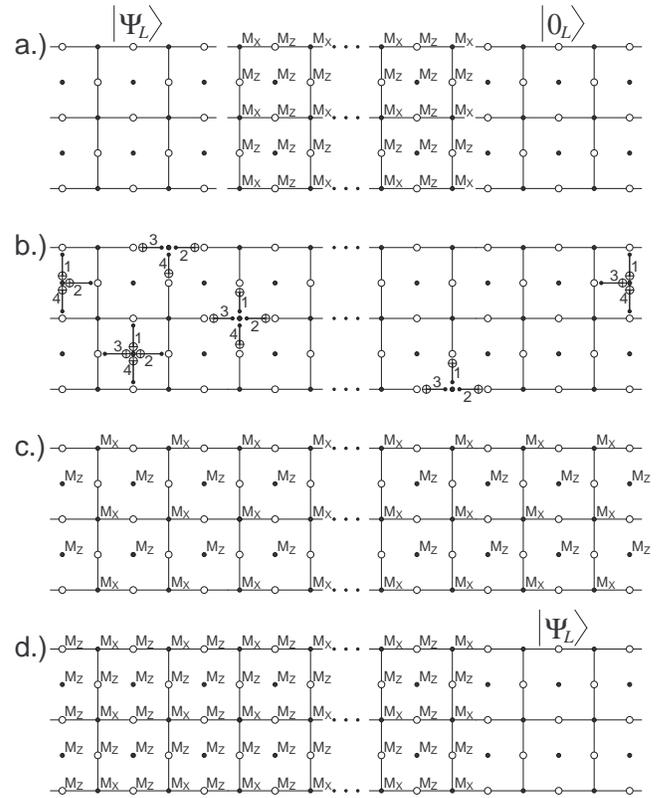}}
\end{center}
\caption{Monolithic surface code quantum communication. a.)
Monolithic lattice of qubits with source logical qubit
$|\Psi_L\rangle$, initial measurement pattern for the intermediate
region, and destination area initialized to $|0_L\rangle$. b.)
Circuits used in parallel to prepare for stabilizer measurement.
Numbers indicate the timing of gates. c.) Intermediate stabilizer
measurements. d.) Final stabilizer measurement and communicated
state.}\label{fig:communication}
\end{figure}

We now describe our communication protocol, initially restricting
ourselves to moving a logical qubit from the left end to the right
end of a single monolithic array of qubits with the ability to
perform local gates. Given an arbitrary surface code logical qubit
$|\Psi_L\rangle$ at the left end of the array, an uninitialized
region of qubits $|\Psi\rangle$ in the middle and a surface code
logical qubit $|0_L\rangle$ at the right end, $|\Psi_L\rangle$ can
be fault-tolerantly teleported to the location of $|0_L\rangle$.
First, the uninitialized region is measured as shown in
Fig.~\ref{fig:communication}a. The $Z$ basis measurements project
the region into eigenstates of the $Z$ stabilizers. Second, the
syndrome qubits across the entire lattice are interacted with their
neighboring data qubits as shown in Fig.~\ref{fig:communication}b.
Third, the measurement pattern shown in
Fig.~\ref{fig:communication}c completes one round of stabilizer
measurement. The interaction pattern of
Fig.~\ref{fig:communication}b is executed a total of $d$ times,
interleaved with the measurement pattern of
Fig.~\ref{fig:communication}c. Finally, after the $d$th round of
interaction, the measurement pattern shown in
Fig.~\ref{fig:communication}d is applied, completing the
fault-tolerant movement of the logical qubit.

All measurement results are simply sent to the destination end of
the lattice, not processed during transmission. The final round of
measurements prepares the lattice for the transmission of the next
logical qubit. Assuming each interacting quantum gate takes $T_g$
seconds and each measurement $T_m$ seconds, a logical qubit can be
transmitted every $(4T_g+T_m)d$ seconds. The scaling of $d$ and
values required for practical communication will be discussed later
after the full communication scheme has been described.

The processing of measurement results related to $X$ and $Z$
stabilizers occurs independently. Errors result in stabilizer
measurements changing. A chain of errors leads to changes in the
stabilizer measurements only at the endpoints of the chain. A good
approximation of the most likely pattern of errors corresponding to
a given set of stabilizer measurement changes is one in which every
change is connected by a chain of errors to another change or
lattice boundary such that the total number of errors is a minimum.
A classical algorithm, the minimum weight perfect matching algorithm
\cite{Cook99}, can find such a pattern efficiently, in time growing
poly-logarithmically with the volume of the lattice when parallel
processing is used \cite{Devi09}. An alternative algorithm with
similar runtime has been devised recently \cite{Guil09}. Error
correction fails when the corrections actually create error chains
connecting pairs of opposing boundaries. With careful calculation of
the distance between changes, a minimum of $\lfloor (d+1)/2 \rfloor$
errors must occur before failure is possible, implying $p_c$
decreases exponentially with $d$.

\begin{figure}
\begin{center}
\resizebox{85mm}{!}{\includegraphics{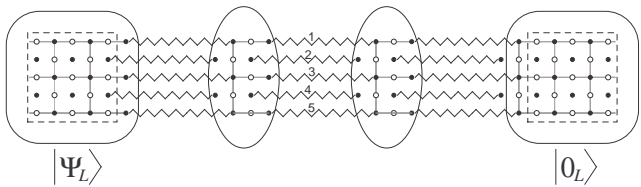}}
\end{center}
\caption{Repeater-based surface code quantum communication. The
qubit pattern in each quantum repeater (ellipses) is for $d=3$. The
pattern width is independent of $d$.}\label{fig:scheme}
\end{figure}

\begin{figure}
\begin{center}
\resizebox{85mm}{!}{\includegraphics{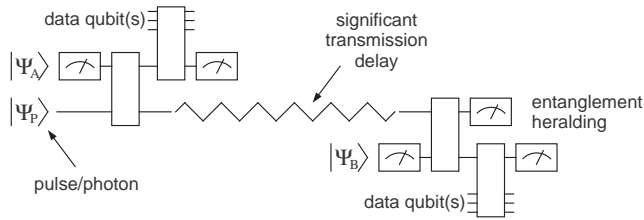}}
\end{center}
\caption{If the probability of heralded success is sufficiently
high, qubit A can be interacted with its neighboring data qubits and
measured before the entangling pulse/photon even reaches its
destination. Error correction takes care of heralded failures,
including loss during transmission.}\label{fig:entangle}
\end{figure}

When communicating over a large physical distance, the fundamental
entanglement resource is expected to be Bell pairs created over
fiber links kilometers in length. The monolithic lattice described
above can be broken into pieces connected by Bell pairs as shown in
Fig.~\ref{fig:scheme}. Stabilizers spanning the communication link
can be measured using the approach shown in Fig.~\ref{fig:entangle}.
We temporarily ignore heralded failure to entangle, which is
discussed below. The left half of each Bell pair can be measured
before the right half even reaches its destination. The rate of the
scheme thus remains unchanged --- one logical qubit every
$(4T_g+T_m)d$ seconds. Latency is, however, introduced as the qubits
in any given repeater station are not initialized until the first
photons arrive from the left. For many ranges of parameters, a given
repeater will have finished working and sending photons before the
next repeater receives its first photons.

The scheme's maximum tolerable Bell pair error rate is of critical
importance. Let us temporarily assume that all gates within repeater
nodes are perfect and Bell pairs are subject to depolarizing errors.
We shall continue to ignore heralded failure to entangle for the
moment. A probability $p_B$ of depolarizing error on a Bell pair
means that the errors $IX$, $IY$, $IZ$, $XI$, $XX$, $XY$, $XZ$,
$YI$, $YX$, $YY$, $YZ$, $ZI$, $ZX$, $ZY$, $ZZ$ each occur with
probability $p_B/15$. Using the Bell pair stabilizers $XX$ and $ZZ$,
these errors are equivalent to $II$ with probability $p_B/5$ and
$IX$, $IY$, $IZ$ with equal probability $4p_B/15$.

After correction, nontrivial combinations of $X$/$Z$ errors form a
chain that runs from the top spatial/temporal boundary to the bottom
spatial/temporal boundary. Given this symmetry, and the fact that
the different types of errors are processed independently, we focus
on $IX$ errors, which occur on any given Bell pair with probability
$p_X=8p_B/15$. Referring to the Bell pairs numbered 1 to $2d-1$ in
Fig.~\ref{fig:scheme}, $IX$ errors on odd pairs induce an $X$ error
on the data qubit to their left whereas on even pairs the result is
an incorrect stabilizer measurement.

These errors can be visualized as the bonds of a $d\times t$ 2-D
square lattice. The error rate $p_X$ is too high when, after
correction, the probability of having a chain of errors along the
$d$ dimension increases with $d$. For $t=1$, we have a repetition
code, implying $p_X<0.5$ is correctable. For $t=d$, we have a
surface code with perfect syndrome measurement implying $p_X\lesssim
0.1$ \cite{Wang09}. The equivalent values of $p_B$ are $15/16$ and
approximately $0.2$.

\begin{figure}
\begin{center}
\resizebox{75mm}{!}{\epsfig{figure=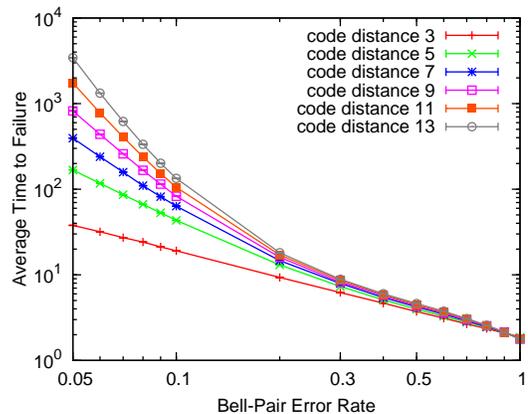, angle=-90}}
\end{center}
\caption{Average number of error correction rounds before logical
failure versus Bell pair error rate $p_B$ and code
distance.}\label{fig:threshold}
\end{figure}

We simulated a pair of repeater nodes with perfect gates and
depolarized Bell pairs for verification (Fig.~\ref{fig:threshold}).
Note the expected crossover at $p_B=15/16\sim 0.94$. Significant
growth of the time to failure with $d$ occurs for $p_B \lesssim
0.2$, as expected. Rapid growth occurs for $p_B\sim 0.1$, equivalent
to a fidelity $F$ of the entangled state $\rho$ with respect to the
desired Bell state $\ket{\Phi^+}$ of 0.92 since
$F=\bra{\Phi^+}\rho\ket{\Phi^+}=1-4p/5$ for Bell pairs corrupted by
depolarizing errors.

\begin{figure}
\begin{center}
\resizebox{75mm}{!}{\epsfig{figure=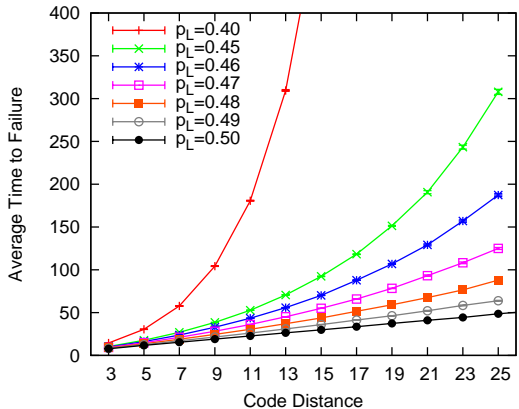, angle=-90}}
\end{center}
\caption{Average number of error correction rounds before logical
failure versus loss and code distance.}\label{fig:qcommloss}
\end{figure}

Loss during transmission can be modeled as measurement in an unknown
basis. Loss is easier to tolerate than depolarizing noise as the
failure to measure the transmitted pulse or photon gives the
location of the error. This can be seen in the simulation results of
Fig.~\ref{fig:qcommloss}, which shows efficient handling of 40-45\%
loss. Note that no code can handle more than 50\% loss as this would
violate the no-cloning theorem \cite{Diek82,Woot82}.

\begin{figure}
\begin{center}
\resizebox{75mm}{!}{\epsfig{figure=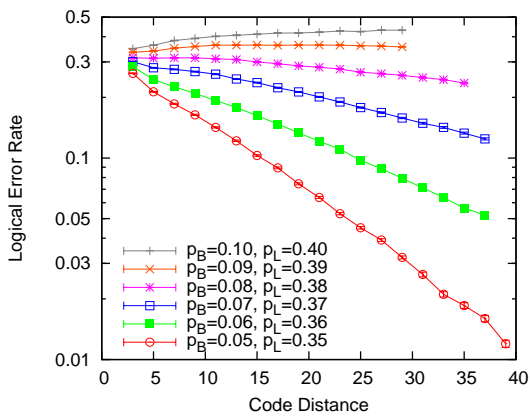, angle=-90}}
\end{center}
\caption{Probability of logical error per link for a variety of loss
and Bell error rates.}\label{fig:bellloss}
\end{figure}

The probability of logical error after $d$ successful stabilizer
measurements, $p_{\rm link}$, is shown in Fig.~\ref{fig:bellloss}
versus $p_B$ and loss $p_L$. For 35\% loss and 5\% error ($F=0.96$),
increasing $d$ by 30 decreases $p_{\rm link}$ by a factor of 10.
Sending data through $10^4$ repeaters with $10^{-6}$ error would
require $d\sim 300$, corresponding to of order a thousand qubits per
repeater. Each repeater takes time $d(4T_g+T_m)/(1-p_L)$ to send a
logical qubit. Long-range, high fidelity MHz communication can thus
be achieved provided $300(4T_g+T_m)/0.65\sim 1$$\mu$s, meaning
$\sim$2ns gates.

Permitting repeaters to have a nonzero local gate error rate $p_g$
will only have significant impact if it is close to the threshold
error rate of approximately $p_g^{th}=0.75\%$ \cite{Raus07}. An
error rate one or two orders of magnitude below this will not
significantly change the above results.

To summarize, we have shown that, provided the Bell pair error rate
is less than approximately 10\% ($F\gtrsim 0.92$), utilizing surface
code quantum error correction enables the practical fault-tolerant
quantum communication of logical qubits over an arbitrary number of
links $N$ with arbitrarily low communication error rate $p_c$ given
$O(\log N/p_c)$ qubits per repeater. If the rate of loss is high,
the communication time is proportional to the time $T_B$ required to
successfully create a Bell pair and the number of Bell pairs per
link $O(\log^2 N/p_c)$. If the loss is below approximately 35\% and
$F\gtrsim 0.96$, no heralding is required and of order a thousand
qubits per repeater and nanosecond gates enables one to send logical
qubits at a MHz rate with $10^{-6}$ error through $10^4$ links ---
sufficient in principle to reach the opposite side of the planet.

We acknowledge helpful discussions with Bill Munro, Simon Devitt,
Ashley Stephens and Sean Barrett. AGF, DSW, CDH, LLCH acknowledge
support from the Australian Research Council, the Australian
Government, and the US National Security Agency (NSA) and the Army
Research Office (ARO) under contract number W911NF-08-1-0527. RV
acknowledges support from JSPS. TDL was partially supported by the
National Science Foundation CCR-08 29694, MEXT, and NICT.

\bibliography{../../../References}

\end{document}